\documentclass[12pt]{article}

\usepackage{a4}

\def\tr{\mathop{\rm tr}}
\def\U{{\rm U}}
\def\SU{{\rm SU}}
\newcommand{\ol}{\overline}
\newcommand{\wt}{\widetilde}

\usepackage{amsmath}
\usepackage{amsfonts}
\usepackage{bm}

\newcommand{\CC}{\mathbb{C}}
\newcommand{\ZZ}{\mathbb{Z}}
\newcommand{\mm}{\mathfrak{m}}

\begin{document}

\begin{titlepage}
\title{\hfill\parbox{4cm}
       {\normalsize UT-09-08\\March 2009}\\
       \vspace{2cm}
Charges and homologies in AdS$_4$/CFT$_3$
       \vspace{2cm}}
\author{
Yosuke Imamura\thanks{E-mail: \tt imamura@hep-th.phys.s.u-tokyo.ac.jp}
\\[30pt]
{\it Department of Physics, University of Tokyo,}\\
{\it Hongo 7-3-1, Bunkyo-ku, Tokyo 113-0033, Japan}
}
\date{}

\maketitle
\thispagestyle{empty}

\vspace{0cm}

\begin{abstract}
\normalsize
Electric and magnetic charges of a certain class of operators
in ${\cal N}=2$ large $N$
quiver Chern-Simons theories are investigated.
We consider only non-chiral theories, in which
every bi-fundamental field
appears with its conjugate representation.
By interpreting operators in a Chern-Simons theory
as wrapped M-branes in the dual geometry $AdS_4\times X_7$,
we partly determine the homologies of $X_7$.
\end{abstract}

\end{titlepage}

\section{Introduction}
AdS$_4$/CFT$_3$ correspondence claims the equivalence
between a three-dimensional
conformal field theory (CFT) and M-theory in the background
\begin{equation}
AdS_4\times X_7,
\end{equation}
with an appropriately chosen Einstein $7$-manifold $X_7$,
which may in general contain singularities.
In the brane picture, the CFT is the theory realized on
multiple M2-branes in the cone over $X_7$.
Until quite recent, low-energy effective field theories realized on
multiple M2-branes were not known.
The first example of interacting CFTs describing multiple M2-branes
were proposed by Bagger and Lambert \cite{Bagger:2006sk,Bagger:2007jr,Bagger:2007vi}
and Gusstavson \cite{Gustavsson:2007vu,Gustavsson:2008dy}.
The model, which is now called BLG model, is a Chern-Simons theory
with ${\cal N}=8$ supersymmetry, and is important also as the first
example of Chern-Simons theories with ${\cal N}\geq4$ supersymmetry.
Although BLG model works only for two M2-branes \cite{Lambert:2008et,Distler:2008mk},
it triggered the following construction
and classification
of various supersymmetric quiver Chern-Simons theories \cite{Gaiotto:2008sd,Fuji:2008yj,Hosomichi:2008jd,Aharony:2008ug,Hosomichi:2008jb,Bagger:2008se,Schnabl:2008wj,Imamura:2008dt}.
${\cal N}\leq3$ quiver Chern-Simons theories, whose actions have
been known for long, are also investigated from new perspective
as theories for
multiple M2-branes in various backgrounds \cite{Ooguri:2008dk,Jafferis:2008qz,Martelli:2008si,Hanany:2008cd,Ueda:2008hx,Imamura:2008qs,Hanany:2008fj,Franco:2008um,Hanany:2008gx}.

Almost all theories appearing in the recent literature as
theories on M2-branes are quiver Chern-Simons theories.%
\footnote{Recently, ${\cal N}=3$ Chern-Simons theories with fundamental flavors
describing multiple M2-branes in hyper-K\"ahler cone backgrounds
were proposed in \cite{Hohenegger:2009as,Gaiotto:2009tk}. We do not discuss such theories in this paper.}
The purpose of this paper is to investigate
the relation between charges in
such quiver Chern-Simons theories and wrapped M-branes in the
corresponding internal space $X_7$.

In the case of AdS$_5$/CFT$_4$ duality,
the relation between wrapped branes and operators are
first proposed in \cite{Witten:1998xy} for non-dynamical (external)
baryonic operators in the maximally
supersymmetric Yang-Mills theories,
and was extended to dynamical baryonic operators
in theories with less supersymmetries
in \cite{Gubser:1998fp,Gukov:1998kn}.
In the latter, baryonic operators are identified
with D3-branes wrapped on three-cycles in internal
five-dimensional spaces.
This is further extended
in \cite{Butti:2006au,Forcella:2007wk,Butti:2007jv}
to a large class of quiver gauge theories
described by brane tilings \cite{Hanany:2005ve,Franco:2005rj,Franco:2005sm}.
In such a class of gauge theories, the
spectrum of baryonic operators is known to be consistent with the
homology of three-cycles, on which D3-branes can wrap.

The relation between fractional branes,
D5-branes wrapped on two-cycles, and
ranks of $\SU(N)$ gauge groups is
also well understood \cite{Klebanov:2000nc,Klebanov:2000hb}
in the case of
Klebanov-Witten theory \cite{Klebanov:1998hh}.
See also \cite{Benvenuti:2004wx,Butti:2006hc} for fractional branes
in more general theories described by brane tilings.

In this paper, we discuss similar relations
for AdS$_4$/CFT$_3$ from the field theory side.
We investigate electric and magnetic charges
of operators, and relate them to wrapping numbers
of M5- and M2-branes.
We determine one-, two-, and five-cycle homologies of $X_7$
by the analysis of charges on the field theory side.
We also discuss the relation between ranks of gauge groups
and the four-form flux in the internal space.

We emphasize that operators considered in this paper
are not always gauge invariant.
In general, they are charged under the $\U(1)$
part of $\U(N)\sim\SU(N)\times\U(1)$ gauge groups.
We do not impose gauge invariance with respect to all these $\U(1)$ subgroups.
This is because, as is pointed out in \cite{Imamura:2008ji},
it is impossible to identify all wrapped M-branes
which are particles in AdS$_4$
with gauge invariant operators.
We require operators to be singlet with respect to $\SU(N)$ subgroups
in the gauge group.
We refer to such $\SU(N)$ invariant operators as
``colorless operators.''

Colorless operators are constructed by contracting all
$\SU(N)$ color indices of constituent objects.
For simplicity, we do not consider the full symmetry structure
of $\SU(N)$ indices.
For example, when an operator has two color indices,
we do not take care about whether the indices are symmetric or anti-symmetric.
We only take account of the ``index number'' $z$ defined by
\begin{equation}
z=
\mbox{\# of upper $\SU(N)$ indices}
-\mbox{\# of lower $\SU(N)$ indices}.
\end{equation}
We determine whether a combination of
operators can be colorless by only checking
whether the total index number vanishes.
Of course, this simple prescription cannot capture the
detailed spectrum of operators,
and more careful analysis is necessary when we want to determine
degeneracy of operators and so on.
We leave this problem for future work.

As we mentioned above, colorless operators are in general charged under
the diagonal $\U(1)$'s of $\U(N)$'s in the gauge group.
We call this baryonic symmetry, and
refer to operators rotated by this symmetry as
baryonic operators.
Such operators are constructed with $\SU(N)$ epsilon
tensors, and expected to have conformal dimension of order $N$.
It is known that the mass of M5-branes wrapped on five cycles
reproduce this scaling of the conformal dimension,
and by this reason, we identify baryonic operators
with wrapped M5-branes.
Once we accept this correspondence,
it is natural to identify
monopole operators, operators magnetically charged with respect to
the baryonic symmetry,
with M2-branes wrapped on two-cycles,
which are mutually non-local to the wrapped M5-branes.

In general, quantum corrections shift the baryonic charges
of monopole operators \cite{Borokhov:2002ib,Borokhov:2002cg},
and they make the charge spectrum of such operators complicated.
Unfortunately,
we have not succeeded in interpreting such a complicated spectrum
in terms of M-branes.
By this reason, in this paper, we discuss only the
non-chiral ${\cal N}=2$ quiver Chern-Simons theories,
in which a bi-fundamental chiral multiplet
in $(N_a,\ol N_b)$
and one in $(\ol N_a,N_b)$ appear in the pairwise way
and the corrections to
baryonic charges vanish.
Examples of such non-chiral theories describing M2-branes
are ${\cal N}=3$ Chern-Simons theories
studied in \cite{Jafferis:2008qz}.

This paper is organized as follows.
In \S\ref{sun.sec} we define the colorless sector
for a single $\U(N)$ gauge group.
In \S\ref{quiver.sec} we generalize this into quiver Chern-Simons theories,
and declare the class of operators we discuss.
In \S\ref{mon.sec}
we relate M2-branes wrapped on two-cycles to non-baryonic
monopole operators.
In \S\ref{bar.sec}, we discuss the correspondence between
baryonic operators and wrapped M5-branes.
In \S\ref{str.sec} we study how flux strings are
realized as wrapped M-branes.
(By flux strings we mean stringy objects in AdS$_4$.
Since we consider CFT, confining strings do not exist
on the field theory side.)
A relation among ranks of gauge groups, baryonic charges,
and the charge of flux strings attached on baryonic operators
are also studied.
In \S\ref{exp.sec} we present some examples.
The last section is devoted to discussions.

\section{$\U(N)$ gauge group}\label{sun.sec}
In this preliminary section,
we consider a gauge theory with a single $\U(N)$ gauge group.
We denote this gauge group by $G$, and the gauge field by $A$.
We assume that $\U(N)$ is the effective gauge group.
If there were no matter fields the gauge group would be
effectively $\SU(N)/\ZZ_N$ because
the diagonal $\U(1)$ would not couple to any fields.
We do not consider such a case and
assume the existence of fields coupled by the full $\U(N)$ gauge group.

We can use Young diagrams to specify $\U(N)$ representations.
Let $w_i$ be the number of boxes in the $i$-th row in a Young diagram.
These numbers form the highest weight vector for
the representation.
It is an element of the $\U(N)$ weight lattice ${\cal W}$.
\begin{equation}
\overrightarrow w=(w_1,w_2,\ldots,w_N)\in{\cal W}.
\end{equation}
When we use a weight vector to specify Young diagram,
the components are ordered in the descending order;
$w_1\geq w_2\geq \cdots\geq w_N$.
Because we consider not $\SU(N)$ but $\U(N)$,
columns filled up with $N$ boxes have meaning and
we cannot neglect them.
In other words, we should distinguish between $(w_1,\ldots,w_N)$
and $(w_1+1,\ldots,w_N+1)$.
Note that negative $w_i$ are not prohibited.
Although the $\SU(N)$ representation and the $\U(1)$ charge
are independent for general $\U(N)$ representations,
we consider only the case in which
both the $\U(1)$ charge and the $\SU(N)$ representation
are specified by the same Young diagram.

We define monopoles in $\U(N)$ gauge theory
following Goddard, Nuyts, and Olive \cite{Goddard:1976qe}.
Namely, we consider Dirac monopoles for the Cartan subgroup $H=\U(1)^N\subset G$.
Such monopoles are characterized by the Dirac strings attached on them,
and their magnetic charges are defined by integrating the gauge potential around
the Dirac string as
\begin{equation}
m_i=\frac{1}{2\pi}\oint A_i,\quad
i=1,\ldots,N,
\end{equation}
where the index $i$ labels the $\U(1)$'s in $H$,
and $A_i$ is $i$-th diagonal component in the $\U(N)$ gauge field $A$.
Monopoles defined in this way are called
Goddard-Nuyts-Olive (GNO) monopoles.
We can regard the set of magnetic charges $m_i$ as a vector
in the root lattice ${\cal R}$:
\begin{equation}
\overrightarrow m=(m_1,m_2,\ldots,m_N)\in{\cal R}.
\end{equation}

The weight lattice ${\cal W}$ and the root lattice ${\cal R}$ are
dual to each other, and
for vectors $\overrightarrow w\in{\cal W}$
and $\overrightarrow m\in{\cal R}$
the inner product
\begin{equation}
\overrightarrow w\cdot\overrightarrow m\equiv\sum_{i=1}^Nw_im_i
\end{equation}
is defined.

We mainly focus on ``the colorless sector''
as is mentioned in Introduction.
This means that we focus on the electric and magnetic charges
with respect to the diagonal $\U(1)$ subgroup of $\U(N)$.
When we discuss electric charge in the colorless sector,
we consider only $\SU(N)$ singlet operators.
$\SU(N)$ non-singlet representations are
excluded from the consideration.
This constrains weight vectors by
``the colorless condition''
\begin{equation}
\overrightarrow w\cdot\overrightarrow \alpha_a=0,\quad
a=1,\ldots,N-1,
\label{coles}
\end{equation}
where $\overrightarrow \alpha_a\in{\cal R}$ are the $\SU(N)$ root vectors defined by
\begin{eqnarray}
\overrightarrow\alpha_1&=&(1,-1,0,\ldots,0),\nonumber\\
\overrightarrow\alpha_2&=&(0,1,-1,0,\ldots,0),\nonumber\\
&\vdots&\nonumber\\
\overrightarrow\alpha_{N-1}&=&(0,\ldots,0,1,-1).
\end{eqnarray}
The colorless condition
(\ref{coles}) means
\begin{equation}
w_1=w_2=\cdots=w_N=:b,
\label{colorless}
\end{equation}
and the Young diagram of a colorless representation is
an $N\times b$ rectangle.

If we consider only colorless operators,
the effective gauge group becomes
\begin{equation}
G_B=\U(N)/\SU(N)=\U(1)/\ZZ_N,
\end{equation}
where $\U(1)$ in the last expression means the diagonal $\U(1)$
subgroup of $\U(N)$ and $\ZZ_N$ is the center of $\SU(N)$.
We define the $G_B$ gauge field by
\begin{equation}
B=\tr A.
\end{equation}
This couples to the fundamental representation by charge $1/N$.

Contrary to the electric charges,
we do not impose any restriction to the magnetic charge $\overrightarrow m$.
Instead, we simply neglect
the magnetic charges other than the $G_B$ charge.
This is realized by introducing the following equivalence relation:
\begin{equation}
\overrightarrow m\sim
\overrightarrow m+\sum_{a=1}^{N-1}c_a\overrightarrow\alpha_a,
\quad
c_a\in\ZZ.
\label{unequiv}
\end{equation}
This identification removes $N-1$ components of the magnetic charges,
and leaves information of the $G_B$ magnetic charge only.

In general, when we consider
a pairing of two linear spaces with inner product between them,
an equivalence relation
in one space
always arises with
a constraint in the other space
for the consistency with the inner product.
In the case of charge lattices we discuss here,
the inner product $\overrightarrow w\cdot\overrightarrow m$
is well-defined in the colorless sector because it
does not depend on
the choice of an element from an equivalence class defined by (\ref{unequiv})
thanks to the restriction (\ref{coles}).

When we consider
the colorless sector, we can use a single integer to
represent each of electric and magnetic charges.
For the electric charge, we use the common value $b$ in (\ref{colorless}),
while an equivalence class of the $G_B$ magnetic charge
defined by (\ref{unequiv})
is specified by
\begin{equation}
m=\sum_{i=1}^Nm_i.
\label{colorlessproj}
\end{equation}
The inner product of electric and magnetic charge vectors
is equal to the product of these integers.
\begin{equation}
\overrightarrow w\cdot\overrightarrow m=bm.
\end{equation}

For concreteness, let us consider $\U(N)$ gauge theory
with chiral multiplets $Q^\alpha$ and $\wt Q_\alpha$
in the fundamental and anti-fundamental representation,
respectively.
We also assume the existence of the Chern-Simons term
\begin{equation}
S_{\rm CS}=\frac{k}{4\pi}\int\tr\left(AdA+\frac{2}{3}A^3\right).
\end{equation}
Colorless operators in this Chern-Simons theory
are constructed by combining the following objects:
\begin{itemize}
\item The component fields in $Q^\alpha$ and $\wt Q_\alpha$.
\item $\SU(N)$ invariant anti-symmetric tensors
$\epsilon_{\alpha_1\cdots\alpha_N}$ and
$\epsilon^{\alpha_1\cdots\alpha_N}$.
\item Monopole operators.
\end{itemize}

If the Chern-Simons level is $k$,
a monopole operator $\mm[\overrightarrow m]$
with magnetic charge $\overrightarrow m$ belongs to
$\SU(N)$ representation specified by the weight vector
$\overrightarrow w=k\overrightarrow m$.
In general, the $G_B$ charge of the operator
receives quantum corrections.
As we mentioned in Introduction,
we consider only non-chiral
theories in which such corrections vanish.
Then the $G_B$ charge is given by $b=km/N$,
and we can regard $\overrightarrow w$ as a $\U(N)$ weight vector.

The above mentioned monopole operators are
elementary ones before combined with matter fields
to form colorless operators.
We use the character $\mm$ to denote such ``bare'' monopole operators.
The index number and the electric and the magnetic charges of
these objects are shown in Table \ref{table:un}.
\begin{table}[htb]
\caption{The index number $z$, the electric $G_B$ charge $b$,
and the magnetic $G_B$ charge $m$ of elementary objects
are shown.}
\label{table:un}
\begin{center}
\begin{tabular}{cccc}
\hline
\hline
    & $z$ & $b$ & $m$ \\
\hline
$Q^\alpha$ & $1$ & $1/N$ & $0$ \\
$\wt Q_\alpha$ & $-1$ & $-1/N$ & $0$ \\
$\epsilon_{\alpha_1\cdots\alpha_N}$ & $-N$ & $0$ & $0$ \\
$\epsilon^{\alpha_1\cdots\alpha_N}$ & $N$ & $0$ & $0$ \\
$\mm[\overrightarrow m]$ & $km$ & $km/N$ & $m$ \\
\hline
\end{tabular}
\end{center}
\end{table}
We construct colorless operators
by combining these objects so that the
index number $z$ cancels.
For such operators, the electric charge $b$ is always
an integer and is the same as
the number of the epsilon tensor.
(We mean by ``the number of the epsilon tensor''
the number of $\epsilon_{\alpha_1\cdots\alpha_N}$
subtracted by that of $\epsilon^{\alpha_1\cdots\alpha_N}$.)
Namely, the charge $b$ counts the number of $\SU(N)$ ``baryons''.
This is the reason why we call $G_B$ the baryonic symmetry.

\section{Quiver Chern-Simons theories}\label{quiver.sec}
Let us extend the arguments in the last section to
quiver gauge theories.
We consider a quiver Chern-Simons theory
described by a connected quiver diagram with $n$ vertices.
The gauge group is given by
\begin{equation}
G=\prod_{a=1}^n \U(N_a),
\label{qgaugeg}
\end{equation}
and the action includes the Chern-Simons terms
\begin{equation}
S_{\rm CS}=\sum_{a=1}^n\frac{k_a}{4\pi}\int\tr\left(A_adA_a+\frac{2}{3}A_a^3\right).
\label{csun}
\end{equation}
We define the ``color part'' of the gauge group by
\begin{equation}
G_{\SU}=\prod_{a=1}^n \SU(N_a)\subset G.
\end{equation}
Note that we do not remove the diagonal $\U(1)$ subgroup of $G$ which
does not act on any matter fields in the gauge theory.
This is because
we implicitly assume that the theory is embedded in
string or M-theory.
In such a case we can introduce an external source belonging to the
fundamental representation in a $\U(N_a)$ gauge group,
to which the diagonal $\U(1)$ subgroup couples.
We later impose a certain condition (eq. (\ref{clcons}))
to exclude such representation
from the physical spectrum.

We consider the colorless sector of this quiver gauge theory.
The baryonic symmetry $G_B$ is defined as the effective group acting on
colorless operators:
\begin{equation}
G_B=G/G_{\SU}
=\prod_{a=1}^n (\U(1)'_a/\ZZ_{N_a})
=\prod_{a=1}^n \U(1)_a,
\end{equation}
where $\U(1)'_a$ is the diagonal subgroup of $\U(N_a)$ and
$\U(1)_a$ is its quotient by $\ZZ_{N_a}$, the center of $\SU(N_a)$.
Let $B_a$ be the $\U(1)_a$ gauge field defined by
\begin{equation}
B_a=\tr A_a.\label{btra}
\end{equation}

For each $\U(N_a)$ factor, we define electric and magnetic charges
of the colorless sector in the same way as the previous section.
We denote the electric and magnetic $\U(1)_a$ charge by $b_a$ and $m_a$,
respectively.
We collect them to form the vectors
\begin{equation}
{\bm b}=(b_1,\ldots,b_n),\quad
{\bm m}=(m_1,\ldots,m_n),
\label{vectexpa}
\end{equation}
and define the inner product
\begin{equation}
{\bm b}\cdot{\bm m}=\sum_{a=1}^nb_am_a.
\end{equation}
Each component of the vectors in (\ref{vectexpa}) corresponds to
each $\U(N_a)$ factor in the gauge group $G$,
or, equivalently, each vertex in the quiver diagram,
while
each component of vectors
$\overrightarrow w$ or $\overrightarrow m$ used in the last section
corresponds to each $\U(1)$ factor in the
Cartan subgroup of a single $\U(N)$.


Colorless operators are constructed by combining
\begin{itemize}
\item Bi-fundamental chiral multiplets $\Phi_I=(\phi_I,\psi_I)$
\item $\SU(N_a)$ invariant anti-symmetric tensors
$\epsilon_{(a)\alpha_1\cdots\alpha_{N_a}}$
and $\epsilon_{(a)}^{\alpha_1\cdots\alpha_{N_a}}$
\item Monopole operators $\mm[{\bm m}]$
\end{itemize}
We define charge matrix $\{Q_{Ia}\}$
so that the component $Q_{Ia}$ is $+1$ ($-1$) if
$\Phi_I$ belongs to the fundamental (anti-fundamental) representation of $\U(N_a)$, and otherwise $Q_{Ia}=0$.

The $\SU(N_a)$ index numbers, the electric and magnetic $\U(1)_a$ charges
of the fundamental objects are shown in
Table \ref{table:quiver}.
\begin{table}[htb]
\caption{The index numbers, the electric and magnetic charges are shown}
\label{table:quiver}
\begin{center}
\begin{tabular}{cccc}
\hline
\hline
    & $z_a$ & $b_a$ & $m_a$ \\
\hline
$\phi_I$, $\psi_I$ & $Q_{Ia}$ & $Q_{Ia}/N_a$ & $0$ \\
$\epsilon_{(a)}$ & $-N_a$ & $0$ & $0$ \\
$\mm[{\bm m}]$ & $k_am_a$ & $k_am_a/N_a$ & $m_a$ \\
\hline
\end{tabular}
\end{center}
\end{table}
We can again easily see that for colorless operators
the electric charge $b_a$ is always
an integer and is the same as the number of $\SU(N_a)$
invariant anti-symmetric tensor $\epsilon_{(a)}$
included in the operator.
Thus we can regard the charge $b_a$ as the $\SU(N_a)$ baryon number.
The complete contraction of color indices
is possible only when the relation
\begin{equation}
N_ab_a-k_am_a=z_a[\Phi]
\label{isq}
\end{equation}
holds, where $z_a[\Phi]$ is the $\SU(N_a)$ index number
carried by bi-fundamental fields in the operator.
Because all matter fields in a quiver gauge theory are
bi-fundamental fields, the right hand side
in (\ref{isq}) vanishes when it is summed up with respect to $a$.
We obtain
\begin{equation}
{\bm N}\cdot{\bm b}-{\bm k}\cdot{\bm m}=0.
\label{clcons}
\end{equation}
We define the charge lattice $\Gamma$ of colorless operators
as the set of vectors
$({\bm b},{\bm m})$ satisfying
(\ref{clcons}).
\begin{equation}
\Gamma=
\{({\bm b},{\bm m})|{\bm N}\cdot{\bm b}-{\bm k}\cdot{\bm m}=0\}
=\ZZ^{2n-1}.
\end{equation}

To relate the charges $({\bm b},{\bm m})$ and wrapping numbers of
M-branes on the gravity side is a main purpose of this paper.
In order to have clear geometric picture with wrapped branes,
we take the large $N$ limit.
We assume that the ranks $N_a$ are given by
\begin{equation}
N_a=N+\delta N_a,
\label{nandn}
\end{equation}
and take the large $N$ limit with $\delta N_a$ fixed at order $1$.
We also assume that the charges $b_a$ and $m_a$ are of order $1$.
\begin{equation}
b_a\sim{\cal O}(1),\quad
m_a\sim{\cal O}(1).
\label{smallcharges}
\end{equation}
If the charges are of order $N$
the corresponding branes would deform the background geometry
and the probe approximation would cease to be valid.
Although it would be very interesting to investigate
such a deformed geometry,
we restrict ourselves to the case in which we can treat
the branes as probes.

We separate
operators into two classes, non-baryonic and baryonic operators.
Non-baryonic operators are defined as operators with $\bm b=0$.
The other operators with $\bm b\neq0$ are referred to as
baryonic operators.
By definition, the non-baryonic operators are not only
colorless but also gauge invariant.
They are in general monopole operators carrying magnetic charges.

\section{Monopoles and two-cycles}\label{mon.sec}
Let us first discuss correspondence between non-baryonic (monopole)
operators and
wrapped M2-branes.
By definition, non-baryonic operators are characterized
by only the magnetic charge $\bm m$
constrained by
\begin{equation}
{\bm k}\cdot{\bm m}=0.
\end{equation}
The vector ${\bm m}$ satisfying this condition spans
the sublattice $\Gamma_M\subset\Gamma$ defined by
\begin{equation}
\Gamma_M=
\{(0,{\bm m})|{\bm k}\cdot{\bm m}=0\}
=\ZZ^{n-1}.
\end{equation}
We would like to relate monopole operators to wrapped M2-branes.
However, it is known that 
a certain subset of these operators
does not correspond to wrapped branes but to bulk Kaluza-Klein modes.

Let us temporarily consider the case with $N_a=1$.
We can regard this Abelian Chern-Simons theory
as a subsector of the non-Abelian theory
representing the motion of a single M2-brane.
In the subsector, the Chern-Simons action (\ref{csun})
reduces to
\begin{equation}
S_{\rm CS}=\sum_{a=1}^n\frac{k_a}{4\pi}\int A_adA_a,
\label{csu1}
\end{equation}
where $A_a$ in this action should be interpreted as
one of diagonal components of $\U(N_a)$ gauge field
corresponding to the single M2-brane we are focusing on.
Let us re-organize the $n$ $\U(1)$ gauge fields $A_a$ into
the diagonal $\U(1)$ gauge field $A_D$ and the other $n-1$ gauge fields
$A_i'$ ($i=1,\ldots,n-1$).
The relation between $A_a$ and $(A_D,A_i')$ is
\begin{equation}
A_a=A_D+A_a'
\end{equation}
where $A_a'$ are linear combinations of $A_i'$.
By substituting this into (\ref{csu1}) we obtain
\begin{equation}
S_{\rm CS}
=\frac{1}{4\pi}\sum_{a=1}^nk_a\int A_DdA_D
+\frac{1}{2\pi}\int A_Dd\sum_{a=1}^nk_aA_a'
+\sum_{a=1}^n\frac{k_a}{4\pi}\int A'_adA'_a.
\label{csexp}
\end{equation}
If we assume
\begin{equation}
{\bm 1}\cdot{\bm k}=0,\quad
{\bm1}\equiv(1,1,\ldots,1),
\label{totkcond}
\end{equation}
then the first term on the right hand side in
(\ref{csexp}) vanishes and the diagonal $\U(1)$
gauge field $A_D$ appears in the action only through the
second term in (\ref{csexp}).
The equation of motion of $A_D$ is
\begin{equation}
d\sum_{a=1}^nk_aA_a=0,
\end{equation}
and we can solve this by
\begin{equation}
\sum_{a=1}^nk_aA_a=da.
\end{equation}
The scalar field $a$ is the dual-photon field.
This is periodic scalar field with period $2\pi$ and
plays a role of the coordinate of the ``eleventh'' direction
in the M-theory background.
In the following, the relation (\ref{totkcond}) is
always assumed because otherwise we cannot regard the theory
as a theory of M2-branes.

Due to the periodicity of $a$, it is natural to define the
operator $e^{ia}$.
Because $a$ is the canonical conjugate to the flux $(2\pi)^{-1}dA_D$,
the operator $e^{ia}$ changes the flux $(2\pi)^{-1}dA_D$ by one.
In other words, it carries the diagonal magnetic charge
\begin{equation}
{\bm m}={\bm 1}.
\label{m11}
\end{equation}
In the non-Abelian quiver gauge theory with gauge group (\ref{qgaugeg}),
we should extend this operator to
the monopole operator $\mm[{\bm1}]$ carrying the magnetic charge
(\ref{m11}).
By combining $\mm[{\bm1}]$ and the matter fields,
we can always make colorless monopole operators with the same
magnetic charge:
\begin{equation}
{\cal O}=\mm[{\bm1}]\prod_I(\phi_I)^{s_I}.
\label{MM}
\end{equation}
The index numbers of $\mm[{\bm1}]$ are $z_a=k_a$, and
for the operator (\ref{MM}) to be colorless,
$s_I$ must be integers solving
the equation
\begin{equation}
k_a+\sum_IQ_{aI}s_I=0\quad
\forall a.
\end{equation}
If $s_I$ is negative, $(\phi_I)^{s_I}$ should be interpreted
as $(\phi_I^\dagger)^{-s_I}$.
Thanks to (\ref{totkcond})
and the connectivity of the quiver diagram,
solutions always exist.
If we would like to obtain chiral operators,
we cannot use $\phi_I^\dagger$ and $s_I$ should be non-negative.
Because we assume the
theory is non-chiral and
the quiver diagram is
not only connected but also strongly connected, namely,
every vertex is reachable
from every other following oriented edges,
the existence of such solutions is guaranteed.

In the correspondence between
non-baryonic operators and wrapped M2-branes,
we should exclude operators whose charges
are multiple of (\ref{m11}).
The exclusion of such operators
is realized by introducing the equivalence relation
\begin{equation}
{\bm m}\sim{\bm m}+{\bm 1},
\end{equation}
in the lattice $\Gamma_M$.
We define the group of magnetic charges
corresponding to wrapped M2-branes by
\begin{equation}
\Gamma_{M2}=\Gamma_M/({\bm m}\sim{\bm m}+{\bm 1})
=\ZZ^{n-2}.
\end{equation}
Note that the constraint ${\bm k}\cdot{\bm m}=0$ and the equivalence
relation ${\bm m}\sim{\bm m}+{\bm1}$ are consistent to each other
thanks to the assumption (\ref{totkcond}).
We identify this group with the two-cycle homology of the
internal space $X_7$:
\begin{equation}
H_2(X_7)=\Gamma_{M2}=\ZZ^{n-2}.
\label{h2}
\end{equation}

\section{Baryons and five-cycles}\label{bar.sec}
Let us consider baryonic operators with ${\bm b}\neq0$.
We do not impose the condition ${\bm m}=0$ for baryonic operators,
and in general baryonic operator may carry
magnetic charges.
We define the group of baryonic charges
by neglecting the charges of monopole operators.
Namely, we define the charge lattice of baryonic operators
as the following quotient lattice:
\begin{equation}
\Gamma_B=\Gamma/\Gamma_M.
\end{equation}
The constraint (\ref{clcons}) gives
\begin{equation}
{\bm N}\cdot{\bm b}=0\mod\gcd{\bm k}.
\label{clcons2}
\end{equation}
In the large $N$ limit,
by using (\ref{nandn}) and (\ref{smallcharges}),
we decompose this
into the conditions
\begin{equation}
{\bm 1}\cdot{\bm b}=0,
\label{clcons2b}
\end{equation}
and
\begin{equation}
\bm{\delta N}\cdot{\bm b}=0\mod\gcd{\bm k}.
\label{clcons2a}
\end{equation}
The first condition (\ref{clcons2b})
guarantees that
(\ref{clcons}) can be satisfied with
$\bm m$ of order $1$.
This condition is necessary because we only consider operators
realized on the gravity side as probe branes.
When (\ref{clcons2b}) is satisfied,
(\ref{clcons2}) becomes the constraint
(\ref{clcons2a}), which will be regarded as
the condition for the absence of flux strings
attached on the operator.

Although the condition
(\ref{clcons2a}) must hold for the operator to be colorless,
it is convenient to define
the lattice $\Gamma_B'$ defined only by the first constraint
(\ref{clcons2b}).
\begin{equation}
\Gamma_B'=\{\bm b\in\ZZ^n|{\bm 1}\cdot{\bm b}=0\}=\ZZ^{n-1}.
\end{equation}
A vector in $\Gamma_B'$ in general gives colored operators accompanied
by flux strings, and the second condition (\ref{clcons2a})
defines the lattice $\Gamma_B$ of colorless operators
as a sublattice of $\Gamma_B'$.

Similarly to the case of non-baryonic operators,
a certain subset of $\Gamma_B$
does not correspond to wrapped M5-branes.
Let us consider
$\bm{\delta N}=0$ case first.
In this case,
we can define the dual-photon field
in the non-Abelian theory in the same way as
the Abelian ($N_a=1$) case.
The dual photon field is defined by
\begin{equation}
da=\sum_{a=1}^nk_aB_a
\end{equation}
where $B_a$, which is defined in (\ref{btra}), is the gauge fields coupling to the
charge $b_a$.
Under gauge transformation $\delta B_a=d\lambda_a$,
the dual photon field is transformed by
\begin{equation}
\delta a=\sum_{a=1}^nk_a\lambda_a.
\end{equation}
Due to this non-linear gauge transformation,
the expectation value of the dual photon field breaks a $\U(1)$ subgroup of $G_B$
into a certain discrete group.
Therefore, the charge associated with this broken $\U(1)$ is no longer conserved,
and cannot be identified with any wrapping number of M5-branes on
the gravity side.
Thus, to remove this unconserved component from the charges,
we introduce the equivalence relation
\begin{equation}
{\bm b}\sim{\bm b}+{\bm k},
\label{unhigged}
\end{equation}
representing the ``screening'' by the operator $e^{ia}$,
and define the baryonic charge group by
\begin{equation}
\Gamma'_{M5}=\Gamma_B'/({\bm b}\sim{\bm b}+{\bm k})
=\ZZ^{n-2}\times \ZZ_{\gcd{\bm k}}.
\label{qbdef}
\end{equation}
Note that if $\bm{\delta N}=0$
(\ref{clcons2a}) is automatically satisfied
and $\Gamma_B'=\Gamma_B$.
We identify the group (\ref{qbdef})
with the five-cycle homology group $H_5(X_7)$.

Let us next consider the general case with $\bm{\delta N}\neq0$.
Even in this case
the topology of the internal space
is expected not to change from
the case of $\bm{\delta N}=0$ as long as $\bm{\delta N}$ is of order one,
and we still identify the group
(\ref{qbdef}) with the five-cycle homology.
\begin{equation}
H_5(X_7)=\Gamma'_{M5}=\ZZ^{n-2}\times \ZZ_{\gcd{\bm k}}.
\label{h5}
\end{equation}
It is, however, not necessarily the same as the group of
isolated wrapped
M5-branes because wrapped M5-branes are in general accompanied by
flux strings realized as wrapped M2-branes
as is studied in more detail in the next section.
We define the group of wrapped M5-branes
without flux strings
as a subset of $\Gamma_{M5}'$ by requiring the condition (\ref{clcons2a}).
\begin{equation}
\Gamma_{M5}=\{[{\bm b}]\in\Gamma_{M5}'|\bm{\delta N}\cdot{\bm b}=0\mod\gcd{\bm k}\},
\label{qbdef2}
\end{equation}
where $[{\bm b}]={\bm b}+\ZZ{\bm k}$ is
the equivalence class with representative ${\bm b}$.
The inner product in (\ref{qbdef2})
as an element of $\ZZ_{\gcd\bm k}$
does not depend
on the choice of a representative from $[{\bm b}]$
because $\bm{\delta N}\cdot{\bm k}=0\mod\gcd{\bm k}$.

The combination of
two conditions (\ref{clcons2b}) and (\ref{clcons2a})
is equivalent to the single condition (\ref{clcons2})
only under the restriction (\ref{smallcharges}).
If we permit magnetic charge of order $N$,
there exist colorless operators whose charge $\bm b$
satisfies (\ref{clcons2}),
but not (\ref{clcons2b}) and (\ref{clcons2a})
separately.
An example of such operators is
the following monopole operator associated with a single $\U(N)$ gauge group:
\begin{equation}
\epsilon_{\alpha_1\cdots\alpha_N}
\mm[\overrightarrow m]^{\alpha_1\cdots\alpha_N},\quad
\overrightarrow m=(1,1,\ldots,1).
\label{epsm}
\end{equation}
This operator, however, is
prohibited by a gauge invariance condition as we explain below.
The reason why we have not imposed gauge invariance
with respect to the $\U(1)$ part of $\U(N)$ groups
is that some of gauge fields of these $\U(1)$  are
regarded as the boundary values
of bulk gauge fields,
and couple to
wrapped M-branes \cite{Imamura:2008ji}.
If such a bulk gauge field is absent for a
$\U(1)$ gauge symmetry on the boundary,
the gauge invariance with respect to this $\U(1)$ must be imposed.
Once we accept the relation (\ref{h5}),
we have only $b_5=n-1$ bulk gauge fields coupling to wrapped M5-branes,
and no bulk gauge field couples
the diagonal baryonic charge ${\bm 1}\cdot{\bm b}$.
Therefore,
concerning this diagonal part,
we must impose the gauge invariance condition,
which is nothing but (\ref{clcons}).
Thus, the operator (\ref{epsm}) does not
have its counterpart on the gravity side.

\section{Flux strings and ranks of gauge groups}\label{str.sec}
In general if we introduce an external source of the color charge,
the charge is partially screened by ambient fields.
Well-known example is that
the color charge in the $\SU(N)$ pure Yang-Mills theory
is screened by the adjoint field to leave only the ``$N$-aliy'' of the representation.
In a confining theory, an external source
with unscreened charge is accompanied by a
flux string.
This is not the case in non-confining theories.
Even in such non-confining theories,
on the gravity side,
operators with unscreened
charge is treated as endpoints of stringy objects
in AdS space.
For example, external quarks in the maximally supersymmetric
Yang-Mills theory in four dimensions are
treated as the endpoints of fundamental strings on the
conformal boundary \cite{Rey:1998ik,Maldacena:1998im}.
We will use the term ``flux strings'' in the following
to mean such stringy objects in AdS$_4$.

In this section
we treat baryonic operators as external sources,
and discuss flux strings attached on them.
What degrees of freedom is left after screening
in a quiver Chern-Simons theory?
If we take account of the vacuum polarization of
adjoint fields, the information of a $\U(N_a)$ representation
is almost lost and we are left with
only the index number $z_a$ for each $\U(N_a)$.
The polarization of bi-fundamental fields hides the distinction
among $\U(N_a)$ factors, and only the total index number
\begin{equation}
z=\sum_{a=1}^nz_a
\end{equation}
is left.
If we take account of all non-baryonic operators
including monopole operators,
only the modulo $\gcd{\bm k}$ part of $z$ is left unscreened
because as is shown in Table \ref{table:quiver}
the index number of monopole operators
is linear combination of Chern-Simons levels $k_a$ with
integral coefficients.
If this unscreened charge does not vanish,
the source is accompanied by
flux strings.

In the previous section, we saw that the electric
charge $\bm b$ of a colorless baryonic operator
satisfies (\ref{clcons2a}).
We can regard this as the condition for
the complete screening of the charges of
the operator.
If (\ref{clcons2a}) does not hold,
the baryonic operator is accompanied by
a flux string with
charge
\begin{equation}
f=\bm{\delta N}\cdot\bm{b}\in\ZZ_{\gcd{\bm k}}.
\label{samen}
\end{equation}

On the gravity side,
we can interpret this relation as follows.
Let us consider a baryonic operator realized as
an M5-brane wrapped on a five-cycle $\Omega_5$.
The action of the M5-brane includes
\begin{equation}
\frac{1}{2\pi}\oint_{M5}H_3\wedge C_3
=\frac{1}{2\pi}\oint_{M5}b_2\wedge F_4,
\label{fcoupling}
\end{equation}
where $H_3=db_2$ is the field strength of the two-form field
$b_2$ living on the M5-brane, and $F_4=dC_3$ is the field strength
of the background three-form field $C_3$.
Let $\Sigma_3\in H_3(X_7)$ be the Poincare dual of
the background four-form flux $[(2\pi)^{-1}F_4]\in H^4(X_7)$.
Through the interaction (\ref{fcoupling}),
the background flux induces the charge
on the M5-brane worldvolume electrically coupled by the field $b_2$.
Because $\Omega_5$ is compact, the charge
must be canceled by the charge of the boundary of M2-branes attached on
the M5-brane.
For this cancellation
we need to attach M2-brane along the one-cycle $\gamma_1$ in $\Omega_5$
which is Poincare dual in $\Omega_5$ to
\begin{equation}
\left[\frac{1}{2\pi}F_4|_{\Omega_5}\right]\in H^4(\Omega_5).
\end{equation}
In other words, $\gamma_1$ is the intersection of
the five-cycle $\Omega_5$ and the three-cycle $\Sigma_3$
\begin{equation}
\gamma_1=\Omega_5\cap\Sigma_3.
\label{gammaos}
\end{equation}
This is the geometric translation of the relation
(\ref{samen}).
We identify flux strings with M2-branes wrapped on one-cycles,
and the flux string charge group with the one-cycle homology
\begin{equation}
H_1(X_7)=\ZZ_{\gcd{\bm k}}.
\label{h1}
\end{equation}
Flux strings generate non-trivial monodromies for
baryonic operators (wrapped M5-branes).

Up to now, we have obtained the following homologies
by the comparison of operators in a Chern-Simons theory
and their M-brane realizations:
\begin{equation}
H_1(X_7)=\ZZ_{\gcd{\bm k}},\quad
H_2(X_7)=\ZZ^{n-2},\quad
H_5(X_7)=\ZZ^{n-2}\times\ZZ_{\gcd{\bm k}}.
\end{equation}
These are consistent
to the duality of the homology groups.
In general, the following duality relations hold
among homologies of $d$-dimensional manifold:
\begin{equation}
H_i^f=H_{d-i}^f,\quad
H_i^t=H_{d-i-1}^t,
\label{duality}
\end{equation}
where $H_i^f$ and $H_i^t$ are the free part and the torsion subgroup,
respectively, of the homology $H_i$.

In the above argument, we relate the three-cycle homology $H_3(X_7)$
to $\bm{\delta N}$, the ``fractional'' part of the ranks:
\begin{equation}
\mbox{three-cycles}\quad
\leftrightarrow\quad
\bm{\delta N}.
\label{threedn}
\end{equation}
In general, the structure of the three-cycle homology is
highly non-trivial, and
we do not try to establish the concrete map between
three-cycles and $\bm{\delta N}$.
We here comment on one important point;
$\bm{\delta N}=0$ does not necessarily mean
vanishing four-form flux.
If all the ranks are the same and $\bm{\delta N}=0$,
(\ref{clcons2a}) is automatically satisfied.
On the gravity side, this means that the three-cycle $\Sigma_3$ satisfies
\begin{equation}
[\Omega_5\cap\Sigma_3]=[0]\in H_1(X_7)
\quad
\forall[\Omega_5]\in H_5(X_7).
\label{equaln}
\end{equation}
The condition (\ref{equaln})
does not require $[\Sigma_3]=0$, the vanishing background
four-form flux.
Indeed, in the case of ${\cal N}=4$
Chern-Simons theories,
there are in general many possible $F_4$ discrete torsion
corresponding to equal-rank quiver Chern-Simons theories \cite{Imamura:2008ji}.
We will mention such an example
in the following section.

In addition to M2-branes wrapped on one-cycles,
there is another potential origin of
stringy objects:
M5-branes wrapped on four-cycles.
Combining the two-cycle homology
(\ref{h2}) and the duality relation (\ref{duality}),
we find that $H_4(X_7)$ does not have torsion subgroup.
The duality relation also says that it is the same as the free
part of $H_3(X_7)$.
\begin{equation}
H_4(X_7)
=H_3^f(X_7)
=\ZZ^{b_3}.
\label{h4}
\end{equation}
Absence of the torsion subgroup in $H_4(X_7)$ means that
associated strings does not induce
fractional monodromies for monopole operators.
We have no idea about interpretation of these strings.
We only comment that these may have something to do with
the cascading phenomenon.
If $b_3\neq0$, we can introduce
four-form flux in the free part of the four-form cohomology
group $H^4(X_7)=H_3(X_7)$.
Unlike the discrete torsion, such a flux induces non-vanishing energy
and deforms the background geometry.
Such a deformation signals the existence of
cascading phenomenon \cite{Klebanov:2000hb}
in three dimensions.

The homologies we obtained up to now
are collected in Table \ref{table:hom}.
They are completely determined by
three integers, $s$, $b_2$, and $b_3$,
and torsion part $T$ of $H_3(X_7)$.
\begin{table}[htb]
\caption{Homologies of $X_7$ conjectured from the charge analysis in
quiver Chern-Simons theories.}
\label{table:hom}
\begin{center}
\begin{tabular}{ccccccccc}
\hline
\hline
& $H_0$ & $H_1$ & $H_2$ & $H_3$ & $H_4$ & $H_5$ & $H_6$ & $H_7$ \\
\hline
free part &  $\ZZ$ & $0$ & $\ZZ^{b_2}$ & $\ZZ^{b_3}$ & $\ZZ^{b_3}$ & $\ZZ^{b_2}$ & $0$ & $\ZZ$ \\
torsion part & $0$ & $\ZZ_s$ & $0$ & $T$ & $0$ & $\ZZ_s$ & $0$ & $0$ \\
\hline
\end{tabular}
\end{center}
\end{table}
The integers $s$ and $b_2$ are given
in terms of parameters in the Chern-Simons theory by
\begin{equation}
s=\gcd{\bm k},\quad
b_2=n-2.
\label{sb2}
\end{equation}

\section{Examples}\label{exp.sec}
In the previous sections, by the analysis of charges in
non-chiral Chern-Simons theories,
we conjectured the homology groups of the dual geometry $X_7$
as Table \ref{table:hom}.
Typical examples of Sasaki-Einstein
manifolds/orbifolds are listed in Table \ref{table:example}.
\begin{table}[htb]
\caption{The data $s$, $b_2$, $b_3$, and $T$
of various Einstein manifolds/orbifolds are shown.
Refer to the indicated references for $T$ of the last four manifolds/orbifolds.
(For $V^{5,2}$ only the fact that $H_3$ is at most torsion is mentioned in \cite{Fabbri:1999hw}.)}
\label{table:example}
\begin{center}
\begin{tabular}{ccccc}
\hline
\hline
$X_7$ & $s$ & $b_2$ & $b_3$ & $T$ \\
\hline
${\bf S}^7/\ZZ_k$\cite{Aharony:2008ug}   & $k$ & $0$ & $0$ & $\ZZ_k$ \\
$Q^{1,1,1}$\cite{Fabbri:1999hw}          & $1$ & $2$ & $0$ & $\ZZ_2$ \\
$M^{1,1,1}$\cite{Fabbri:1999hw}          & $1$ & $1$ & $0$ & $\ZZ_9$ \\
$N^{0,1,0}$\cite{Fabbri:1999hw}          & $1$ & $1$ & $0$ & $0$ \\
$V^{5,2}$\cite{Fabbri:1999hw}            & $1$ & $0$ & $0$ & $*$ \\
$Y^{p,k}(\CC P^2)$\cite{Martelli:2008rt} & $\gcd(p,k)$ & $1$ & $0$ & $*$ \\
$Y^{p,k}(\CC P^1\times\CC P^1)$\cite{Martelli:2008rt}
                                         & $\gcd(p,k)$ & $2$ & $0$ & $*$ \\
$({\bf S}^7/(\ZZ_p\times\ZZ_q))/\ZZ_k$\cite{Imamura:2008ji}
                                         & $k$ & $p+q-2$ & $0$ & $*$ \\
\hline
\end{tabular}
\end{center}
\end{table}
All examples in the table have vanishing $b_3$.
It is also known that every $3$-Sasakian manifold has $b_3=0$.
This is not the case for general Einstein manifolds.
The simplest example is ${\bf S}^3\times{\bf S}^4$
with appropriate radii.
Sasaki-Einstein manifolds with $b_3\neq0$ are also known
to exist.
(See \cite{Boyer:1998sf} and references therein.)
Although all manifolds/orbifolds in the table have homologies
consistent with the form
shown in Table \ref{table:hom},
only the first and the last examples
correspond to non-chiral theories,
and
our arguments are not applicable to the others.

In the following subsections, we discuss the two examples of non-chiral
theories in more detail.

\subsection{ABJM model}
The simplest example is the ${\cal N}=6$ Chern-Simons
theory proposed by Aharony, Bergman, Jafferis, and Maldacena \cite{Aharony:2008ug}.
This model, ABJM model,
is described by a quiver diagram with two vertices.
Namely, the gauge group is
$\U(N_1)\times\U(N_2)$.
Due to the condition (\ref{totkcond})
two $\U(N)$ gauge groups have opposite Chern-Simons levels.
$n$ and ${\bm k}$ are given by
\begin{equation}
n=2,\quad
{\bm k}=(k,-k).
\label{abjmdata}
\end{equation}
The Higgs branch moduli space of this theory is the symmetric product of
$\CC^4/\ZZ_k$, and the internal space is $X_7={\bf S}^7/\ZZ_k$.
(\ref{abjmdata}) is consistent
through (\ref{sb2}) with the data of this internal space
shown in Table \ref{table:example}.
The non-trivial homologies are
\begin{equation}
H_0=H_7=\ZZ,\quad
H_1=H_3=H_5=\ZZ_k.
\end{equation}
Let $\sigma_3$ and $\sigma_5$ be the generators of $H_3$ and $H_5$,
respectively.
We can adopt $\sigma_1=\sigma_3\cap\sigma_5$,
the intersection of $\sigma_3$ and $\sigma_5$,
as the generator of the one-cycle homology group.

There are no non-trivial two-cycles in $X_7={\bf S}^7/\ZZ_k$.
Monopole operators in
ABJM model \cite{Berenstein:2008dc,Hosomichi:2008ip,Klebanov:2008vq}
carry ``diagonal'' magnetic charge proportional to ${\bm1}=(1,1)$,
and should be identified with Kaluza-Klein modes in the bulk.

Baryonic operators in ABJM model with $\bm{\delta N}=0$
are studied in \cite{Park:2008bk},
and the degeneracy and the conformal dimension
are reproduced by the analysis
using wrapped branes.
(In \cite{Park:2008bk} the wrapped branes are analyzed
from the perspective of type IIA theory.)

The gravity dual of $\U(N)\times\U(N+M)$ ABJM model is
studied in \cite{Aharony:2008gk}.
The rank difference $M$ in this case
correspond to the fractional brane wrapped on $\Sigma_3=M\sigma_3$,
or, equivalently,
the discrete torsion $F_4=\Sigma_3^*$,
where $\Sigma_3^*$ is the Poincare dual of $\Sigma_3$.
In \cite{Aharony:2008gk} it is argued that
only when $-k\leq M\leq k$ the Chern-Simons theory is unitary,
and two theories with $M=M_1$ and $M=M_2$ are
equivalent if $M_1=M_2\mod k$.

When $0\leq M\leq k$, we can construct
the following baryonic operator:
\begin{equation}
{\cal B}^{\beta_{N+1}\cdots\beta_{N+M}}
=
\epsilon_{\alpha_1\cdots\alpha_N}
\epsilon^{\beta_1\cdots\beta_N\beta_{N+1}\cdots\beta_{N+M}}
\phi^{\alpha_1}_{\beta_1}
\phi^{\alpha_2}_{\beta_2}
\cdots
\phi^{\alpha_N}_{\beta_N},
\label{coloredb}
\end{equation}
where
$\alpha_i$ and $\beta_i$ are $\SU(N)$ and $\SU(N+M)$ indices,
respectively, and
$\phi^\alpha_\beta$ is bi-fundamental scalar field.
ABJM model includes four such bi-fundamental scalar fields.
In (\ref{coloredb}) we omitted the flavor indices
for distinction of these four.
The operator
(\ref{coloredb})
correspond to an M5-brane wrapped on the five-cycle $\sigma_5$.
If $1\leq M\leq k-1$, this is not colorless,
and is accompanied by a flux string with charge $M\in\ZZ_k$.
Geometric description (\ref{gammaos}) of this fact is
\begin{equation}
\Sigma_3\cap\sigma_5=M\sigma_1.
\label{m351}
\end{equation}

If $M=0$, (\ref{coloredb}) is colorless,
and not accompanied by a flux string.
If two theories with $M=0$ and $M=k$ are equivalent
as is argued in \cite{Aharony:2008gk},
it should be possible to construct colorless
baryonic operator even when $M=k$.
Actually, this is possible.
By adding fermionic bi-fundamental fields $\psi^\alpha_\beta$
and the monopole operator $\mm[(-1,0)]$ to (\ref{coloredb}),
we can write the colorless operator
\begin{equation}
{\cal B}
=
\epsilon_{\alpha_1\cdots\alpha_N}
\mm[(-1,0)]_{\alpha_{N+1}\cdots\alpha_{N+k}}
\epsilon^{\beta_1\cdots\beta_N\beta_{N+1}\cdots\beta_{N+k}}
\phi^{\alpha_1}_{\beta_1}
\phi^{\alpha_2}_{\beta_2}
\cdots
\phi^{\alpha_N}_{\beta_N}
\psi_{\beta_{N+1}}^{\alpha_{N+1}}\cdots
\psi_{\beta_{N+k}}^{\alpha_{N+k}}.
\end{equation}

\subsection{${\cal N}=4$ Chern-Simons theories}
${\cal N}=4$ Chern-Simons theories are
described by circular quiver diagrams
whose vertices and edges represent $\U(N)$ gauge groups
and hyper multiplets, respectively.
There are two types of hypermultiplets,
so-called untwisted and twisted hypermultiplets \cite{Hosomichi:2008jd}.
Let $n$, $p$, and $q$ be the number of
vector, untwisted hyper, and twisted hypermultiplets,
respectively.
Because the quiver diagram is circular, the following relation holds:
\begin{equation}
p+q=n.
\end{equation}
The requirement of ${\cal N}=4$ supersymmetry
restricts the Chern-Simons levels to be $\pm k$ or $0$.
The Higgs branch moduli space of this theory is
derived in \cite{Imamura:2008nn}.
See also \cite{Benna:2008zy,Terashima:2008ba}.
It is the symmetric product of
\begin{equation}
(\CC^2/\ZZ_p\times\CC^2/\ZZ_q)/\ZZ_k.
\end{equation}
Correspondingly, the internal space is
\begin{equation}
X_7=({\bf S}^7/(\ZZ_p\times\ZZ_q))/\ZZ_k.
\end{equation}
The data of the homology groups of this orbifold are
given in Table \ref{table:example},
and satisfy the relation (\ref{sb2}).
In \cite{Imamura:2008ji}
not only
the isomorphisms
$\Gamma_{M2}=H_2(X_7)$ and $\Gamma_{M5}'=H_5(X_7)$
but also the agreement of degeneracy and the conformal dimension
of baryonic operators with the predictions of
the M5-brane description is confirmed.
Concerning monopole operators,
the R-charge spectrum are computed on the field theory side
in \cite{Imamura:2009ur}
by using the radial quantization method \cite{Borokhov:2002ib,Borokhov:2002cg},
and agreement with the analysis on the gravity side
is partially confirmed.

An interesting feature of these theories is the non-trivial
structure of $H_3(X_7)$.
It is given by
\begin{equation}
H_3(X_7)=(\ZZ_{kp}^{q-1}\times\ZZ_{kq}^{p-1}\times\ZZ_{kpq})/(\ZZ_p\times\ZZ_q).
\end{equation}
Refer to \cite{Imamura:2008ji} for detailed description of this
homology group.
To understand the meaning of $H_3(X_7)$ on the
field theory side,
it is convenient to realize the theory by type IIB brane system.
When $k=1$, the theory is realized on a system consisting of
D3-branes wrapped around ${\bf S}^1$,
on which gauge theory lives,
and
$p$ NS5 and $q$ D5-branes
intersecting with the D3-branes.
The fivebranes divide the ${\bf S}^1$ into $n$ intervals.
Here, we discuss only the case with $p=q=2$ and $k=1$ for simplicity and concreteness.
In this case, the non-trivial homologies are
\begin{equation}
H_0=H_7=\ZZ,\quad
H_2=H_5=\ZZ^2,\quad
H_3=\ZZ_4.
\end{equation}
In order to specify the brane configuration,
we need to specify the arrangement of the four fivebranes.
For this purpose, we decorate the rank vector in the following way:
\begin{equation}
{\bm N}=(^{1:NS}N_1,^{2:NS}N_2,^{3:D}N_3,^{4:D}N_4{}^{1:NS}).
\label{fbvect}
\end{equation}
The superscripts in (\ref{fbvect})
mean that the fivebranes are arranged along ${\bf S}^1$
in order NS5, NS5, D5, and D5.
The gauge group $\U(N_i)$ corresponding to
$i$-th component of the vector ${\bm N}$ is realized on $N_i$ D3-branes
stretched between two fivebranes indicated
on the two sides of the component $N_i$ in the vector (\ref{fbvect}).
For the brane configuration represented in (\ref{fbvect}),
the levels are
\begin{equation}
{\bm k}=(0,1,0,-1).
\end{equation}
The Chern-Simons level of each $\U(N)$ depends on
the fivebranes at the two ends of the interval,
and we can read off the rule to determine Chern-Simons levels
for general ordering of fivebranes from this example.

Let $\sigma_3$ be the generator of $H_3=\ZZ_4$.
The relation between the $F_4$ discrete torsion and the structure of
the brane system is investigated in \cite{Imamura:2008ji}.
We can also obtain some information about this relation from
the analysis of the monopole spectrum in \cite{Imamura:2009ur}.
Results in these references indicate that
the rank vector of the Chern-Simons theory
corresponding to the discrete torsion $F_4=M\sigma_3^*$ is
\begin{equation}
{\bm N}=(^{1:NS}N,^{2:NS}N+M,^{3:D}N,^{4:D}N^{1:NS}).
\label{nnmnn}
\end{equation}
This is only one of infinitely
many possible choice of the brane configuration,
which are transformed to one another by
continuous interchanges of fivebranes.
Such deformations
are expected to have something to do with
Seiberg-like duality
in three dimensions \cite{Giveon:2008zn,Niarchos:2008jb}.

The rank vector (\ref{nnmnn}) may seem to show that
the equal-rank gauge group is realized
only when the discrete torsion vanishes.
This is, however, not a precise statement
because even if $M\neq0$ it may be possible to realize equal ranks
by interchanges of fivebranes.
Actually, it is possible when $M=0,\pm 1\mod 4$.
In the case of $M=1$,
we can realize equal ranks by exchanging the fivebranes $2$ and $3$.
\begin{equation}
(^{1:NS}N,^{2:NS}N+1,^{3:D}N,^{4:D}N^{1:NS})\stackrel{[23]}{\longrightarrow}
(^{1:NS}N,^{2:D}N,^{3:NS}N,^{4:D}N^{1:NS}).
\end{equation}
We took account of the brane creation due to
the Hanany-Witten effect \cite{Hanany:1996ie}.
The Chern-Simons level for the resulting brane configuration
is ${\bm k}=(1,-1,1,-1)$.
When $M=-1$, we can realize equal ranks by three steps as follows.
\begin{eqnarray}
&&(^{1:NS}N,^{2:NS}N-1,^{3:D}N,^{4:D}N^{1:NS})
\nonumber\\
&&\stackrel{[12]}{\longrightarrow}
\stackrel{[34]}{\longrightarrow}
\stackrel{[14]}{\longrightarrow}
(^{1:D}N-1,^{2:NS}N-1,^{3:D}N-1,^{4:NS}N-1^{1:D}).
\end{eqnarray}
The levels for the resulting brane configuration
is ${\bm k}=(-1,1,-1,1)$.
Similar deformation to equal rank configuration is always possible
if $M=0,\pm1\mod4$.
See \cite{Imamura:2008ji} for detailed analysis of
such brane interchange processes.
If the gauge group is equal-rank,
the condition (\ref{clcons2a}) is trivially satisfied,
and we can define a colorless baryonic operator
corresponding to any five-cycle in $X_7$.

If we start from the rank vector (\ref{nnmnn}) with $M=2\mod 4$,
we cannot arrive at any equal-rank configuration.
Even in this case,
the condition (\ref{clcons2a}) still holds
with $k=1$ and
it should be possible to construct
a colorless baryonic operator for an arbitrary five-cycle.
Let us consider, for example,
a baryonic operator
with charge ${\bm b}=(-1,1,0,0)$ in the theory with
\begin{equation}
{\bm N}=(^{1:NS}N,^{2:NS}N+2,^{3:D}N,^{4:D}N^{1:NS}),\quad
{\bm k}=(0,1,0,-1).
\end{equation}
We can indeed construct the colorless operator
\begin{equation}
{\cal B}=
\epsilon^{\alpha_1\cdots\alpha_N}
\mm^{\beta_{N+1}\beta_{N+2}}
\epsilon_{\beta_1\cdots\beta_N\beta_{N+1}\beta_{N+2}}
\phi_{\alpha_1}^{\beta_1}
\phi_{\alpha_2}^{\beta_2}
\cdots
\phi_{\alpha_N}^{\beta_N},
\label{operb}
\end{equation}
where
$\alpha_i$ and $\beta_i$
are $\U(N_1)$, and $\U(N_2)$ color indices,
respectively.
$\mm$ is a monopole operator with magnetic charge
$(0,2,0,0)$.

\section{Discussions}\label{disc.sec}
In this paper we considered
following aspects in
non-chiral ${\cal N}=2$ quiver Chern-Simons theories:
\begin{itemize}
\item
We defined the lattice $\Gamma_{M2}$ of magnetic charges
of non-baryonic operators,
and identified it with the two-cycle homology $H_2(X_7)$
of the internal space $X_7$.
\item
The lattice of baryonic charge $\Gamma_{M5}'$ were defined.
The colorless baryonic operators forms the sublattice $\Gamma_{M5}\subset\Gamma_{M5}'$.
The former was identified with
the five-cycle homology $H_5(X_7)$.
\item
The charge of flux strings $\ZZ_{\gcd{\bm k}}$ were
identified with the one-cycle homology $H_1(X_7)$.
\item
The charge of flux strings attached on baryonic operators
depends on the ranks of gauge groups,
and is obtained by the relation (\ref{samen}).
We derived the corresponding relation (\ref{gammaos})
on the gravity side by requiring the flux conservation on M5-branes.
We can use these relation to obtain some information
about the relation between ranks of gauge groups and
the four-form flux in the dual geometry.
\end{itemize}

There are many problems which we did not study in
this paper.

The one-to-one correspondence
between operators and dual objects
in AdS$_5$/CFT$_4$ has been intensively investigated.
In particular, in the maximally supersymmetric Yang-Mills theory in four-dimensions,
the duality between
$1/2$ BPS operators classified by
Schur polynomials \cite{Corley:2001zk,Berenstein:2004kk}
and giant gravitons \cite{McGreevy:2000cw,Hashimoto:2000zp}
or bubbling geometries \cite{Lin:2004nb} was found.
It is interesting problem to establish a similar one-to-one
correspondence between operators and objects in the dual geometry
in the case of AdS$_4$/CFT$_3$.

The three-cycle homology group $H_3(X_7)$ is expected to
relate to the rank distribution in the gauge group.
In general, the structure of $H_3$ is complicated,
and it is not straightforward to establish the map
between $\bm{\delta N}$ and elements of $H_3$.
The relation (\ref{gammaos}), which corresponds to
(\ref{samen}) in the field theory,
gives some information.
This is, however, not sufficient to
establish the complete map.
More detailed information may be obtained
by analyzing the spectrum of monopole operators.
Because the three-form potential couples to M2-branes,
it works as Wilson lines shifting the
Kaluza-Klein spectrum of wrapped M2-branes.
By comparing spectrum of monopole operators and
that of wrapped M2-branes we can obtain
additional information about the relation (\ref{threedn}).
Spectrum of monopole operators is studied in \cite{Imamura:2009ur}
for ${\cal N}=4$ Abelian Chern-Simons theories.
Extension of this analysis to more general theories
is important task.

Generalization of our results to chiral theories
is very interesting and challenging problem.
It would enable us to consider a much
larger class of Chern-Simons theories
and Sasaki-Einstein dual geometries, such as
dual pairs constructed
by utilizing brane crystals \cite{Lee:2007kv,Lee:2006hw,Kim:2007ic},
and might provide information about dynamics of Chern-Simons theories
with large quantum corrections.

We hope to return to these problems in the near future.

\section*{Acknowledgements}
I would like to thank Y.~Yasui for valuable information
about Einstein manifolds.
I also would like to thank S.~Yokoyama for discussions.
This work was supported in part by
Grant-in-Aid for Young Scientists (B) (\#19740122) from the Japan
Ministry of Education, Culture, Sports,
Science and Technology.



\begin{thebibliography}{99}
\bibitem{Bagger:2006sk}
  J.~Bagger and N.~Lambert,
  Phys.\ Rev.\  D {\bf 75}, 045020 (2007)
  [arXiv:hep-th/0611108].
\bibitem{Bagger:2007jr}
  J.~Bagger and N.~Lambert,
  Phys.\ Rev.\  D {\bf 77}, 065008 (2008)
  [arXiv:0711.0955 [hep-th]].
\bibitem{Bagger:2007vi}
  J.~Bagger and N.~Lambert,
  JHEP {\bf 0802}, 105 (2008)
  [arXiv:0712.3738 [hep-th]].
\bibitem{Gustavsson:2007vu}
  A.~Gustavsson,
  arXiv:0709.1260 [hep-th].
\bibitem{Gustavsson:2008dy}
  A.~Gustavsson,
  JHEP {\bf 0804}, 083 (2008)
  [arXiv:0802.3456 [hep-th]].
\bibitem{Lambert:2008et}
  N.~Lambert and D.~Tong,
  Phys.\ Rev.\ Lett.\  {\bf 101}, 041602 (2008)
  [arXiv:0804.1114 [hep-th]].
\bibitem{Distler:2008mk}
  J.~Distler, S.~Mukhi, C.~Papageorgakis and M.~Van Raamsdonk,
  JHEP {\bf 0805}, 038 (2008)
  [arXiv:0804.1256 [hep-th]].
\bibitem{Gaiotto:2008sd}
  D.~Gaiotto and E.~Witten,
  arXiv:0804.2907 [hep-th].
\bibitem{Fuji:2008yj}
  H.~Fuji, S.~Terashima and M.~Yamazaki,
  Nucl.\ Phys.\  B {\bf 810}, 354 (2009)
  [arXiv:0805.1997 [hep-th]].
\bibitem{Hosomichi:2008jd}
  K.~Hosomichi, K.~M.~Lee, S.~Lee, S.~Lee and J.~Park,
  JHEP {\bf 0807}, 091 (2008)
  [arXiv:0805.3662 [hep-th]].
\bibitem{Aharony:2008ug}
  O.~Aharony, O.~Bergman, D.~L.~Jafferis and J.~Maldacena,
  arXiv:0806.1218 [hep-th].
\bibitem{Hosomichi:2008jb}
  K.~Hosomichi, K.~M.~Lee, S.~Lee, S.~Lee and J.~Park,
  JHEP {\bf 0809}, 002 (2008)
  [arXiv:0806.4977 [hep-th]].
\bibitem{Bagger:2008se}
  J.~Bagger and N.~Lambert,
  arXiv:0807.0163 [hep-th].
\bibitem{Schnabl:2008wj}
  M.~Schnabl and Y.~Tachikawa,
  arXiv:0807.1102 [hep-th].
\bibitem{Imamura:2008dt}
  Y.~Imamura and K.~Kimura,
    J. High Energy Phys. {\bf10} (2008) 040,
  arXiv:0807.2144.
\bibitem{Ooguri:2008dk}
  H.~Ooguri and C.~S.~Park,
  JHEP {\bf 0811}, 082 (2008)
  [arXiv:0808.0500 [hep-th]].
\bibitem{Jafferis:2008qz}
  D.~L.~Jafferis and A.~Tomasiello,
  JHEP {\bf 0810}, 101 (2008)
  [arXiv:0808.0864 [hep-th]].
\bibitem{Martelli:2008si}
  D.~Martelli and J.~Sparks,
  arXiv:0808.0912 [hep-th].
\bibitem{Hanany:2008cd}
  A.~Hanany and A.~Zaffaroni,
  arXiv:0808.1244 [hep-th].
\bibitem{Ueda:2008hx}
  K.~Ueda and M.~Yamazaki,
  arXiv:0808.3768 [hep-th].
\bibitem{Imamura:2008qs}
  Y.~Imamura and K.~Kimura,
  JHEP {\bf 0810}, 114 (2008)
  [arXiv:0808.4155 [hep-th]].
\bibitem{Hanany:2008fj}
  A.~Hanany, D.~Vegh and A.~Zaffaroni,
  arXiv:0809.1440 [hep-th].
\bibitem{Franco:2008um}
  S.~Franco, A.~Hanany, J.~Park and D.~Rodriguez-Gomez,
  JHEP {\bf 0812}, 110 (2008)
  [arXiv:0809.3237 [hep-th]].
\bibitem{Hanany:2008gx}
  A.~Hanany and Y.~H.~He,
  arXiv:0811.4044 [hep-th].
\bibitem{Hohenegger:2009as}
  S.~Hohenegger and I.~Kirsch,
  arXiv:0903.1730 [hep-th].
\bibitem{Gaiotto:2009tk}
  D.~Gaiotto and D.~L.~Jafferis,
  arXiv:0903.2175 [hep-th].
\bibitem{Witten:1998xy}
  E.~Witten,
  JHEP {\bf 9807}, 006 (1998)
  [arXiv:hep-th/9805112].
\bibitem{Gubser:1998fp}
  S.~S.~Gubser and I.~R.~Klebanov,
  Phys.\ Rev.\  D {\bf 58}, 125025 (1998)
  [arXiv:hep-th/9808075].
\bibitem{Gukov:1998kn}
  S.~Gukov, M.~Rangamani and E.~Witten,
  JHEP {\bf 9812}, 025 (1998)
  [arXiv:hep-th/9811048].
\bibitem{Butti:2006au}
  A.~Butti, D.~Forcella and A.~Zaffaroni,
  JHEP {\bf 0706}, 069 (2007)
  [arXiv:hep-th/0611229].
\bibitem{Forcella:2007wk}
  D.~Forcella, A.~Hanany and A.~Zaffaroni,
  JHEP {\bf 0712}, 022 (2007)
  [arXiv:hep-th/0701236].
\bibitem{Butti:2007jv}
  A.~Butti, D.~Forcella, A.~Hanany, D.~Vegh and A.~Zaffaroni,
  JHEP {\bf 0711}, 092 (2007)
  [arXiv:0705.2771 [hep-th]].
\bibitem{Hanany:2005ve}
  A.~Hanany and K.~D.~Kennaway,
  ``Dimer models and toric diagrams,''
  arXiv:hep-th/0503149.
\bibitem{Franco:2005rj}
  S.~Franco, A.~Hanany, K.~D.~Kennaway, D.~Vegh and B.~Wecht,
  JHEP {\bf 0601} (2006) 096,
  arXiv:hep-th/0504110.
\bibitem{Franco:2005sm}
  S.~Franco, A.~Hanany, D.~Martelli, J.~Sparks, D.~Vegh and B.~Wecht,
  JHEP {\bf 0601} (2006) 128,
  arXiv:hep-th/0505211.
\bibitem{Klebanov:2000nc}
  I.~R.~Klebanov and A.~A.~Tseytlin,
  Nucl.\ Phys.\  B {\bf 578}, 123 (2000)
  [arXiv:hep-th/0002159].
\bibitem{Klebanov:2000hb}
  I.~R.~Klebanov and M.~J.~Strassler,
  JHEP {\bf 0008}, 052 (2000)
  [arXiv:hep-th/0007191].
\bibitem{Klebanov:1998hh}
  I.~R.~Klebanov and E.~Witten,
  Nucl.\ Phys.\  B {\bf 536}, 199 (1998)
  [arXiv:hep-th/9807080].
\bibitem{Benvenuti:2004wx}
  S.~Benvenuti, A.~Hanany and P.~Kazakopoulos,
  JHEP {\bf 0507}, 021 (2005)
  [arXiv:hep-th/0412279].
\bibitem{Butti:2006hc}
  A.~Butti,
  JHEP {\bf 0610}, 080 (2006)
  [arXiv:hep-th/0603253].
\bibitem{Imamura:2008ji}
  Y.~Imamura and S.~Yokoyama,
  arXiv:0812.1331(v3) [hep-th].
\bibitem{Borokhov:2002ib}
  V.~Borokhov, A.~Kapustin and X.~k.~Wu,
  JHEP {\bf 0211}, 049 (2002)
  [arXiv:hep-th/0206054].
\bibitem{Borokhov:2002cg}
  V.~Borokhov, A.~Kapustin and X.~k.~Wu,
  JHEP {\bf 0212}, 044 (2002)
  [arXiv:hep-th/0207074].
\bibitem{Goddard:1976qe}
  P.~Goddard, J.~Nuyts and D.~I.~Olive,
  ``Gauge Theories And Magnetic Charge,''
  Nucl.\ Phys.\  B {\bf 125}, 1 (1977).
\bibitem{Rey:1998ik}
  S.~J.~Rey and J.~T.~Yee,
  Eur.\ Phys.\ J.\  C {\bf 22}, 379 (2001)
  [arXiv:hep-th/9803001].
\bibitem{Maldacena:1998im}
  J.~M.~Maldacena,
  Phys.\ Rev.\ Lett.\  {\bf 80}, 4859 (1998)
  [arXiv:hep-th/9803002].
\bibitem{Fabbri:1999hw}
  D.~Fabbri, P.~Fre', L.~Gualtieri, C.~Reina, A.~Tomasiello, A.~Zaffaroni and A.~Zampa,
  Nucl.\ Phys.\  B {\bf 577}, 547 (2000)
  [arXiv:hep-th/9907219].
\bibitem{Martelli:2008rt}
  D.~Martelli and J.~Sparks,
  JHEP {\bf 0811}, 016 (2008)
  [arXiv:0808.0904 [hep-th]].
\bibitem{Boyer:1998sf}
  C.~P.~Boyer and K.~Galicki,
  Surveys Diff.\ Geom.\  {\bf 7}, 123 (1999)
  [arXiv:hep-th/9810250].
\bibitem{Berenstein:2008dc}
  D.~Berenstein and D.~Trancanelli,
  Phys.\ Rev.\  D {\bf 78}, 106009 (2008)
  [arXiv:0808.2503 [hep-th]].
\bibitem{Hosomichi:2008ip}
  K.~Hosomichi, K.~M.~Lee, S.~Lee, S.~Lee, J.~Park and P.~Yi,
  JHEP {\bf 0811}, 058 (2008)
  [arXiv:0809.1771 [hep-th]].
\bibitem{Klebanov:2008vq}
  I.~Klebanov, T.~Klose and A.~Murugan,
  arXiv:0809.3773 [hep-th].
\bibitem{Park:2008bk}
  C.~S.~Park,
  arXiv:0810.1075 [hep-th].
\bibitem{Aharony:2008gk}
  O.~Aharony, O.~Bergman and D.~L.~Jafferis,
  JHEP {\bf 0811}, 043 (2008)
  [arXiv:0807.4924 [hep-th]].
\bibitem{Imamura:2008nn}
  Y.~Imamura and K.~Kimura,
  Prog. Theor. Phys. {\bf120} (2008) 509,
  arXiv:0806.3727 [hep-th].
\bibitem{Benna:2008zy}
  M.~Benna, I.~Klebanov, T.~Klose and M.~Smedback,
  arXiv:0806.1519 [hep-th].
\bibitem{Terashima:2008ba}
  S.~Terashima and F.~Yagi,
  arXiv:0807.0368 [hep-th].
\bibitem{Imamura:2009ur}
  Y.~Imamura,
  arXiv:0902.4173 [hep-th].
\bibitem{Giveon:2008zn}
  A.~Giveon and D.~Kutasov,
  Nucl.\ Phys.\  B {\bf 812}, 1 (2009)
  [arXiv:0808.0360 [hep-th]].
\bibitem{Niarchos:2008jb}
  V.~Niarchos,
  JHEP {\bf 0811}, 001 (2008)
  [arXiv:0808.2771 [hep-th]].
\bibitem{Hanany:1996ie}
  A.~Hanany and E.~Witten,
  Nucl.\ Phys.\  B {\bf 492}, 152 (1997)
  [arXiv:hep-th/9611230].
\bibitem{Corley:2001zk}
  S.~Corley, A.~Jevicki and S.~Ramgoolam,
  Adv.\ Theor.\ Math.\ Phys.\  {\bf 5}, 809 (2002)
  [arXiv:hep-th/0111222].
\bibitem{Berenstein:2004kk}
  D.~Berenstein,
  JHEP {\bf 0407}, 018 (2004)
  [arXiv:hep-th/0403110].
\bibitem{McGreevy:2000cw}
  J.~McGreevy, L.~Susskind and N.~Toumbas,
  JHEP {\bf 0006}, 008 (2000)
  [arXiv:hep-th/0003075].
\bibitem{Hashimoto:2000zp}
  A.~Hashimoto, S.~Hirano and N.~Itzhaki,
  JHEP {\bf 0008}, 051 (2000)
  [arXiv:hep-th/0008016].
\bibitem{Lin:2004nb}
  H.~Lin, O.~Lunin and J.~M.~Maldacena,
  JHEP {\bf 0410}, 025 (2004)
  [arXiv:hep-th/0409174].
\bibitem{Lee:2007kv}
  S.~Lee, S.~Lee and J.~Park,
  JHEP {\bf 0705}, 004 (2007)
  [arXiv:hep-th/0702120].
\bibitem{Lee:2006hw}
  S.~Lee,
  Phys.\ Rev.\  D {\bf 75}, 101901 (2007)
  [arXiv:hep-th/0610204].
\bibitem{Kim:2007ic}
  S.~Kim, S.~Lee, S.~Lee and J.~Park,
  Nucl.\ Phys.\  B {\bf 797}, 340 (2008)
  [arXiv:0705.3540 [hep-th]].
\end{thebibliography}
\end{document}